# EFFECTIVE COHERENCE LENGTH ESTIMATION OF OPTICAL WAVEFRONTS

Akondi Vyas[1, 2], M B Roopashree[1], B R Prasad[1]

[1]Indian institute of Astrophysics, 2[nd] Block, Koramangala, Bangalore
[2]Indian institute of Science, Bangalore
vyas@iiap.res.in, roopashree@iiap.res.in, brp@iiap.res.in

**Abstract:** In adaptive optics, the measurement of spatial coherence length helps in deciding the optimum design parameters of a Shack Hartmann Sensor (SHS). Two methods of estimating the spatial coherence length of optical wavefronts are presented. The first method is based on counting the number of Hough peaks in the wavefront. The second method is based on a simple data mining technique applied on the wavefronts. Optical wavefronts with different properties are simulated and used for statistical analysis. A comparison of the performance of the two methods is presented using Monte Carlo simulations. It is shown that both these methods can become efficient tools in estimating the effective coherence length of optical wavefronts.

## 1. INTRODUCTION

Adaptive Optics (AO) is a reasonably developed technology used in many areas like astronomical imaging, retinal imaging and free space communication where there is a need for improvement in image quality degraded by randomly fluctuating turbulence along the optical path [1-3].

AO systems sense the incoming optical wavefronts and correct them in real time by imposing a conjugate wavefront over the distorted one [4]. A wavefront sensor, generally a Shack Hartmann Sensor (SHS) is used for sensing wavefronts and a deformable mirror is used for compensating distortions. Wavefront reconstruction algorithm converts data from the SHS into a wavefront. SHS is a two dimensional array of lenses called lenslets. The local gradients of the wavefronts over each lenslet are detected by measuring the shift in the spots near the focal plane of the lenslets. The number of lenslets along the rows and columns decide the resolution of the SHS. The two major factors that need attention during AO corrections are spatial scales of turbulence and temporal scales of fluctuations. In the case of astronomical adaptive optics, keeping an eye over these parameters becomes even more essential since these parameters can vary over a period of time [5].

Continuous monitoring of the spatial coherence helps in deciding the optimum design parameters of the SHS, wavefront reconstruction algorithm and deformable mirror [6]. Using an adaptive SHS, it is possible to dynamically change the number of apertures and hence improve the reconstruction accuracy and consistency for best performance of the real time system [7]. The simplest of wavefront reconstruction algorithms is the least square method operated on the SHS type configuration [8]. AO system performance is enhanced when the resolution of the SHS matches with the spatial coherence scales of turbulence [1]. Turbulence can be visualized as a series of wavefronts each consisting of many smaller locally planar regions (LPRs). These LPRs can be of different sizes and shapes. Monte Carlo simulations were performed on wavefronts with changing coherence length at different SHS resolution. The steps involved in these simulations include generation of the spot pattern (at the focal plane of SHS) due to the incoming wavefronts, reconstruction of the wavefronts using least square reconstructor and finally the calculation of the extent of correlation between the actual wavefront and reconstructed one in terms of the correlation coefficient. Number of LPRs or LPR resolution controls the coherence length and hence is used to evaluate the performance of the SHS. Fig. 1 illustrates that changing coherence length affects the wavefront reconstruction accuracy. It can be seen that the performance of the AO system is best when the LPR resolution matches the sensor resolution.

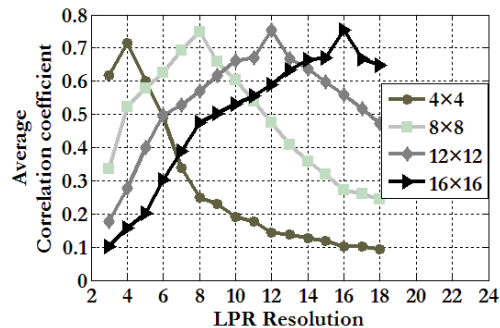

Fig. 1. Performance of AO system as a function of coherence length at different SHS resolutions - 4×4, 8×8, 12×12 and 16×16.

In this paper, an attempt has been made to dynamically estimate the optical coherence length of wavefronts through image processing techniques. Two methods have been proposed in the paper, one of them is based on peak detection using Hough transform technique and another one is based on mining image pixel data.

In the first method, the number of peaks in the image is used as a probe to evaluate the coherence length. The edges in the wavefronts are enhanced. A Hough transform of the edge enhanced image is then used to identify the number of peaks. The peaks above a threshold are counted as valid peaks.

The second method is a more direct one, where data mining is applied on individual optical images. The slope along rows and columns of the wavefront is tracked and the number of times the slope changes its sign is counted. The number of sign changes along rows and columns acts like a probe to estimate effective coherence length.

These methods were tested on wavefronts having structure functions following Kolmogorov spectrum. The simulated wavefronts are stored as two dimensional matrices containing values that represent the magnitudes of fluctuations from the mean and are called phase screens. It is shown by Monte Carlo simulations that both these methods can prove to be effective empirical tools to evaluate optical coherence length and hence enhance the performance of the SHS. The advantages and disadvantages of these methods are illustrated.

## 2. BACKGROUND

### 2.1 Simulation of Phase Screens

Zernike polynomials are orthogonal set of basis functions defined over the unit disk. They can be used to represent a two dimensional function via the computation of Zernike moments [9]. Phase screens following Kolmogorov spatial statistics can be generated by calculating corresponding Zernike moments using the relation derived by Noll [10, 11],

$$
\begin{aligned}
<a_i a_j> = &\ 0.0072 \left(\frac{d}{r_0}\right)^{5/3} (-1)^{(n_i + n_j - 2m_i)/2} \pi^{8/3} \\
&\ \delta_{m_i m_j} \Gamma[(n_i + n_j - 5/3)/2]\{(n_i + 1)(n_j + 1)\}^{1/2} \\
&\ \{\Gamma[(n_i - n_j + 17/3)/2]\ \Gamma[(n_j - n_i + 17/3)/2] \\
&\ \Gamma[(n_i + n_j + 23/3)/2]\}^{-1} \Gamma(14/3)
\end{aligned}
$$

$$(1)$$

where, $a_i$'s represent Zernike moments, d is the telescope aperture diameter and $r_0$ is the Fried parameter, $n$ and $m$ signify radial and azimuthal indices of Zernike polynomials.

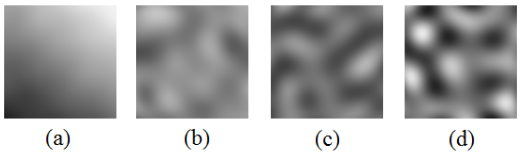

Fig. 2. Simulated Phase Screens using Zernike moment orders (a) 1-180 (b) 20-280 (c) 40-180 (d) 60-180

To simulate phase screens of varying coherence length, the number of moments used to generate them is adjusted. Smaller coherence length phase screens are simulated by excluding the lower order Zernike moments. The phase screens simulated using Zernike moments are shown in Fig. 2.

### 2.2 Structure Function and calculation of $r_0$

Phase fluctuations in the Kolmogorov model of atmospheric turbulence follow a structure function,

$$
D(r) = \left\langle \left| \varphi(\rho + r) - \varphi(\rho) \right|^2 \right\rangle_\rho
$$

$$(2)$$

where $\varphi$ symbolizes the wavefront phase and $D$ stands for the phase variance between two parts of the wavefront separated by '$r$'. $<...>_\rho$ represents ensemble average over '$\rho$'. '$\rho$' represents a vector in the plane perpendicular to optic axis of the telescope. The structure function can also be represented by the Fried parameter $r_0$,

$$
D(r) \propto \left(\frac{|r|}{r_0}\right)^{5/3}
$$

$$(3)$$

From eq. 3, it can be seen that a parameter equivalent to $r_0$ called effective Fried parameter can be easily evaluated by plotting $Deq = D^{3/5}$ against the radial coordinate $|r|$ which follows a straight line as shown in Fig 3. $D(r)$ for the phase screen is evaluated using eq. 2. The radial coordinate, $r$ is normalized such that the largest distance between any two pixels on the phase screen is 1.

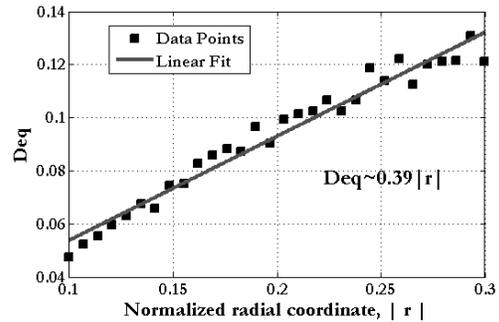

Fig. 3. Estimation of effective Fried parameter for simulated phase screen following Kolmogorov spatial statistics

## 3. METHOD OF HOUGH PEAKS

Hough transform is a pattern recognition technique that picks up local maxima [12]. Multiple phase screens of same coherence length were simulated. The edge enhancement of images will help in better detection of the Hough peaks. The average number of Hough peaks/lines above a threshold is counted by evaluating the Hough transformed matrix for the edge enhanced image of the phase screen. This process is repeated for different coherence length wavefronts.

The number of peaks was found to be increasing linearly with the inverse square of coherence length as shown in Fig. 4.

A linear fit between the square of the inverse of effective coherence length ($R_l$) and the number of peaks (P) gave the following equation,

$$P = 1.3243 \ (R_l)^2 + 17.153 \qquad (4)$$

Fitting a straight line between P and $(R_l)^2$ gives a residual error of 12 peaks when Monte Carlo simulations were performed over 1000 randomly generated phase screens with same coherence length.

Using the expression in eq. 4, it is possible to estimate the coherence length by knowing the number of peaks. Hence, in this method the coherence length has to be estimated by counting the average number of peaks in the Hough transformed matrix of the edge enhanced image.

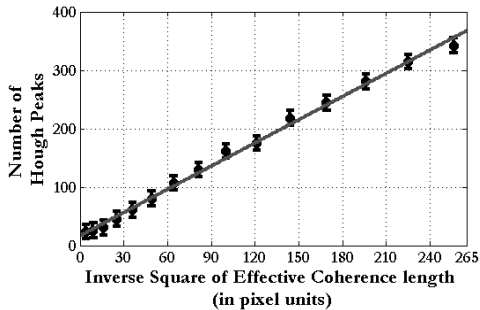

Fig. 4. Measurement of effective coherence length using Hough peak detection

## 4. DATA MINING TECHNIQUE

Data mining is a knowledge discovery technology from large data sets [13]. A logical sequence of steps was followed to estimate the coherence length using this tool. Any optical phase screen can be treated like a 2D matrix or image. A set of pixel values across the rows or columns of the image forms a vector. A difference vector that consists of the adjacent differences has the information of the coherence length in terms of the number of times the sign has changed from positive to negative or vice versa.

The number of sign changes in a single row or column gives the number of crests and troughs in that row or column. This number of sign changes was counted in all the rows and columns and was averaged. This exercise was repeated for multiple phase screens of same coherence length. This process was repeated at different coherence lengths also. The number of sign changes (S) along the row or column varies linearly with inverse of coherence length ($R_l$) as shown in Fig. 5 following the expression,

$$S = 0.70233 \ R_l + 0.72831 \qquad (5)$$

Linearization error in this case is tiny. Following a similar argument made in the last section, $r_0$ can be estimated from the measurement of S, which can be done by taking a large set of wavefronts.

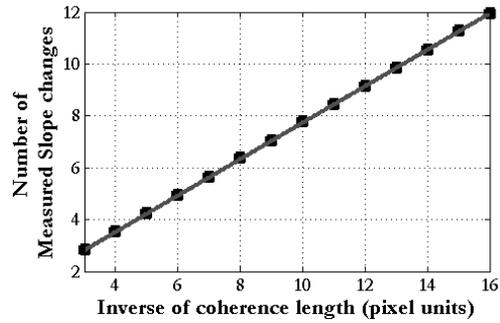

Fig. 5. Effective coherence length measurement using data mining technique

## 5. RESULTS AND DISCUSSION

The coherence length was measured in terms of the image pixels. The estimation error is ±0.5 pixels for Hough peaks method and ±0.1 for data mining procedure when 100 phase screens of similar coherence were used. Error was found to grow with the coherence length in either case. It was also found that increasing the number of samples reduces the error. Optimizing the number of past phase screens required to calculate the coherence length within error thresholds is a problem of significant interest and will be addressed in future.

For accurate measurement in the method involving data mining, a smoother phase screen is needed. Rougher surfaces that are highly irregular have wavefront errors about zero with very small spatial scales and hence counting unnecessary sign changes lead to large coherence length estimation errors. We indicate that the coherence length must be larger than six times the pixel size in the optical phase image recorded by the sensor for distinct recognition.

The choice of threshold in Hough peaks method is critical. Eq. (4) changes with changing threshold. Optimizing the threshold value is essential in the case of Hough peaks. It is most favorable to use the half intensity value as the threshold for peak detection in Hough transformed image.

## 6. CONCLUSION

Two numerical techniques (Hough peak detection and data mining) based on image processing are suggested to estimate the effective coherence length of wavefronts. The measurement of spatial coherence length improves the performance of the AO systems. Monte Carlo simulations were performed on simulated Kolmogorov phase screens to validate these methods. These simulations show that both these methods are effective tools for real time estimation of optical coherence length in adaptive optics. The expressions for coherence length estimation, eq. (4) and eq. (5) should not be taken as a standard, but these empirical laws need to be time-tested and re-examined. The parameters in the expressions should be evaluated using real data.